\documentclass[nonacm]{acmart}
\usepackage{setspace}
\AtBeginDocument{%
  \providecommand\BibTeX{{%
    \normalfont B\kern-0.5em{\scshape i\kern-0.25em b}\kern-0.8em\TeX}}}

\setcopyright{acmcopyright}
\copyrightyear{2018}
\acmYear{2018}
\acmDOI{XXXXXXX.XXXXXXX}

\acmConference[Conference acronym 'XX]{Make sure to enter the correct
  conference title from your rights confirmation emai}{June 03--05,
  2018}{Woodstock, NY}
\acmBooktitle{Woodstock '18: ACM Symposium on Neural Gaze Detection,
 June 03--05, 2018, Woodstock, NY} 
\acmPrice{15.00}
\acmISBN{978-1-4503-XXXX-X/18/06}

\begin{document}

\title{Thinking Upstream: Ethics and Policy Opportunities in AI Supply Chains}


\author{David Gray Widder}
\affiliation{%
  \institution{Carnegie Mellon University}
  \city{Pittsburgh}
  \state{PA}
  \country{USA}
}

\author{Richmond Y. Wong}
\affiliation{%
 \institution{Georgia Tech}
  \city{Atlanta}
 \state{GA}
 \country{USA}}






\maketitle

\section{Looking upstream}
After children were pictured sewing its running shoes in the early 1990s, Nike at first disavowed the ``working conditions in its suppliers’ factories'', before public pressure led them to take responsibility for ethics in their upstream supply chain~\cite{doorey2011transparent}. 
In 2023, OpenAI responded to criticism that Kenyan workers were paid less than \$2 per hour to filter traumatic content from its ChatGPT model by stating in part that it had outsourced the work to a subcontractor, who managed workers' payment and mental health concerns~\cite{2023exclusive}.

In this position paper,\textbf{ we argue that policy interventions for AI Ethics must consider AI as a supply chain problem}, given how the political economy and intra-firm relations structure AI production, in particular examining opportunities \textbf{upstream}. 

Much like physical goods, software is assembled from components developed by many people across diverse contexts, in a ``supply chain''.
Widder and Nafus suggested that ``thinking about ethics and responsibility as chains of relations surfaces specific locations in which ethical decision-making can take place''~\cite{widder2022dislocateda}. They show how interviewees at down the AI supply chain view their ethical responsibility differently:  developers ``high''  in the chain saw what they were making as too ``general purpose'' to warrant ethical scrutiny, but those downstream felt unequipped to remediate ``ethical debt'' accrued upstream.
By analogy to technical debt, ``ethical debt''~\cite{ethicaldebt2020fiesler} refers to unaddressed flaws that may cause harm downstream, to be ``paid'' by developers or users who later interact with the system.

However, AI ethics approaches often focus on the component being developed or its downstream effects, rather than its upstream supply chain. Company AI Ethics policy statements often scrutinize design while avoiding scrutiny of downstream business uses~\cite{greene2019better}. Ethical design principles and checklists are insufficient--- satire shows that systems satisfying ``Fair'', ``Accountable'', ``Transparent'' design principles can nonetheless be patently unethical when \textit{used} for harm~\cite{keyes2019mulching}, and others demonstrate that in cases such as Deepfakes, harm is inherent in how a system is freely distributed and thus widely \textit{used}, not how it is designed~\cite{widder2022limits}. 

Relatedly, efforts like ``Privacy by Design'' and ``Data Protection by Design'' are legal (in some jurisdictions) and engineering requirements that embed privacy concerns throughout the development lifecycle for a particular software component \cite{wong2019bringing}. However, a supply chain frame would consider how that component depends on privacy assumptions or affordances from its dependencies developed earlier in the supply chain, and in turn for its downstream users. G\"{u}rses and Hoboken noted that privacy research and policy interventions focus on sites of ``technology \textit{consumption}'' lower in the supply chain, disregarding modern and drastic changes in software \textit{production}, upstream in the supply chain~\cite{gurses2018agile}.

Ethical design interventions for AI often think downstream, often drawing on design futuring \cite{ballard_judgment_2019,wong2021tactics,fiesler_black_2018,martelaro2020could}, scenarios \cite{zevenbergen2020explainability}, or value sensitive design techniques \cite{shen_value_2021} to consider how stakeholder harms might occur during the deployment and use of AI systems. While useful, we argue that there are unexplored opportunities for acting \textit{upstream}.

\section{Acting upstream}
Conceiving of AI ethics as a supply chain problem, and then looking \textit{up} the chain, surfaces ``values levers''---practices that can pry open discussion about values and ethics \cite{shilton_values_2013} --- that present opportunities for policy, design and activism.

\noindent\textbf{AI Supply Chains and Human Rights Law}. 
Fukuda-Parr and Gibbons argue that government and civil society must act to ground AI ethics in human rights frameworks, but note that company AI ethics guidelines misuse ```human rights' as a rhetorical device'' in the ``absence of enforceable standards'' and regulations~\cite{fukuda-parr2021emerging}. Looking beyond AI supply chains to those of physical goods, such standards and regulations exist. For example, the United States banned ``any goods, wares, articles, and merchandise mined, produced, or manufactured wholly or \textit{in part}'' (emphasis added) in Xinjiang, unless companies can prove they were not produced using Uyghur forced labor~\cite{flacks2022uyghur}, and the UK Modern Slavery and Human Trafficking Act requires companies to prevent forced labor in their supply chain~\cite{kriebitz2020xinjiang}. Considering the working conditions of upstream AI data workers, such as low paid annotators ~\cite{2023exclusive}, suggests opportunities to apply human rights law to workers in the AI supply chain.

\noindent\textbf{Market-Based Policy Interventions: Disclosures, Procurement, and Choosy Customers}. Policy interventions focused on making producers of AI systems disclose information about their upstream practices may create market pressures to address ethical issues. A growing number of institutional investors have expressed interest in investing in companies that meet particular social or ethical standards, often termed “environmental and social good” (ESG) \cite{Schreck2013Disclosure,Kaissar2022Institutional}. While much ESG interest originates in sustainability, monitoring agencies have begun to attend to companies' possible digital harms, including labor rights, data privacy, and security \cite{GlobalReportingInitiativeGRI}. This disclosure is often voluntary and thus non-standardized, but future policy initiatives may explore standardized and compulsory ESG disclosures, incorporating AI supply chain ethics topics. Existing regulatory agencies that require companies to make public disclosures about their business practices, such as the U.S. Securities and Exchange Commission, may play a role here \cite{wong2023privacy}. 

Widder and Nafus explore the power of being a ``choosy'' customer, suggesting that those sourcing software `might routinize asking suppliers for model cards, if the data it was trained on was properly consented, if crowd workers labeling the data were paid an appropriate wage'', as is often done in supply chains for physical goods \cite{widder2022dislocateda}.
While not all customers are choosy, governments can be perhaps more easily required to be so, especially important in cases where governmental uses of AI affect people's freedoms and life chances.
This may also make this scrutiny more routine and thereby normalized. Writing about how government procurement and adoption of ML systems are \textit{policy} decisions as well as technical ones, Mulligan and Bamberger advocate for using the power of procurement to require suppliers to utilize ``contestable design" which ``exposes value-laden features''~\cite{mulligan2019procurement}, which might look like making it possible to challenge upstream software features before their downstream consequences are `baked in'.
Disclosure-based approaches have drawbacks: Gansky and Mcdonald critique ``metadata maximalism'', questioning whether model cards and other ``provenance and trajectory documentation [provided] in order to enable transparency'' can ``steer supplier practice via the discipline of the market''~\cite{gansky2022counterfacctual}, noting inherently messy AI supply chains. 



\noindent\textbf{Design and Activism in the Supply Chain}. 
Design and activist practices can make use of an upstream perspective by creating artifacts and tools that help stakeholders understand, question, and advocate for changes upstream in the AI supply chain. The Algorithmic Equity Toolkit (AEKit) was designed to help citizens and community groups ``find out more about a specific automated decision system'', providing questions for advocates to ask policymakers and technology vendors \cite{Krafft2021toolkit}, including upstream questions like where a system's data came from, who gathered it, with what tools, and for what purpose. Artistic approaches to understanding AI systems, such as Crawford and Joler's ``Anatomy of an AI System'' can help surface and publicize supply chain relationships in AI systems for further questioning \cite{crawford2018anatomy}.  

\noindent\textbf{Ethical Licensing}. Novel software licenses are increasingly proposed and experimented with in open source projects~\cite{contractor2022behavioral,carlosmunozferrandis2022openrail,declaration2018montreal},  which recognize harms from making powerful AI freely available~\cite{widder2022limits}.  As opposed to permissive licenses, these require downstream users to think about their upstream dependencies, and the ethical commitments they demand, and how these fit with their own.

In sum, these ways of acting ``upstream'' present future opportunities for design and policy interventions to address questions of AI Ethics.  





\bibliographystyle{ACM-Reference-Format}
\bibliography{refs}

\end{document}